| TÜRK TARIM ve DOĞA BİLİMLERİ DERGİSİ | TTDB www.dergipark.gov.tr/turkjans | TURKISH JOURNAL of AGRICULTURAL and NATURAL SCIENCES |

Original Article

# Exploring the Impacts of Economic Growth on Ecosystem and Its Subcomponents in Türkiye

Emre Akusta [1]

[1]Kırklareli University, Faculty of Economics and Administrative Sciences, Department of Economics, Kırklareli, Türkiye

[1]https://orcid.org/0000-0002-6147-5443

✉: emre.akusta@klu.edu.tr

**ABSTRACT**

This study analyzes the impacts of economic growth on ecosystem in Türkiye. The study uses annual data for the period 1995-2021 and the ARDL method. The study utilizes the Ecosystem Vitality Index, a sub-dimension of the Environmental Performance Index. In addition, seven models were constructed to assess in detail the impact of economic growth on different dimensions of the ecosystem. The results show that economic growth has a significant impact in all models analyzed. However, the direction of this impact differs across ecosystem components. Economic growth is found to have a positive impact on agriculture and water resources. In these models, a 1% increase in GDP increases the agriculture and water resources indices by 0.074-0.672%. In contrast, economic growth has a negative impact on biodiversity and habitat, ecosystem services, fisheries, acid rain and total ecosystem vitality. In these models, a 1% increase in GDP reduces the indices of biodiversity and habitat, ecosystem services, fisheries, acid rain and total ecosystem vitality by 0.101-2.144%. The results suggest that the environmental costs of economic growth processes need to be considered. Environmentally friendly policies should be combined with sustainable development strategies to reduce the negative impacts of economic growth.

**Key words:** *Economic Growth, Ecosystem, Ecosystem Vitality, Environmental Performance, EPI.*

## Türkiye'de Ekonomik Büyümenin Ekosistem ve Alt Bileşenleri Üzerindeki Etkilerinin Araştırılması

**ÖZ**

Bu çalışma, Türkiye'de ekonomik büyümenin ekosistem üzerindeki etkilerini analiz etmektedir. Çalışmada 1995-2021 dönemi için yıllık veriler ve ARDL yöntemi kullanılmıştır. Çalışmada Çevresel Performans Endeksinin alt boyutu olan Ekosistem Canlılığı endeksi kullanılmıştır. Ayrıca, ekonomik büyümenin ekosistemin farklı boyutları üzerindeki etkisini ayrıntılı olarak değerlendirmek için yedi model oluşturulmuştur. Sonuçlar, ekonomik büyümenin analiz edilen tüm modellerde önemli bir etkiye sahip olduğunu göstermektedir. Ancak bu etkinin yönü ekosistem bileşenleri arasında farklılık göstermektedir. Ekonomik büyümenin tarım ve su kaynakları üzerinde olumlu bir etkisi olduğu tespit edilmiştir. Bu modellerde, GSYH'deki %1'lik bir artış tarım ve su kaynakları endekslerini %0,074-0,672 oranında artırmaktadır. Buna karşılık, ekonomik büyüme biyoçeşitlilik ve habitat, ekosistem hizmetleri, balıkçılık, asit yağmuru ve toplam ekosistem canlılığı üzerinde olumsuz bir etkiye sahiptir. Bu modellerde, GSYH'deki %1'lik bir artış biyoçeşitlilik ve habitat, ekosistem hizmetleri, balıkçılık, asit yağmuru ve toplam ekosistem canlılığı endekslerini %0,101-2,144 oranında azaltmaktadır. Bu da ekonomik büyüme süreçlerinin çevresel maliyetlerinin göz önünde bulundurulması gerektiğini göstermektedir. Ekonomik büyümenin olumsuz etkilerini azaltmak için çevre dostu politikalar sürdürülebilir kalkınma stratejileri ile birleştirilmelidir.

**Anahtar kelimeler:** Ekonomik Büyüme, Ekosistem, Ekosistem Canlılığı, Çevresel Performans, ÇPE.

## INTRODUCTION

The environmental impacts of economic growth have become an increasingly important and complex topic of debate in today's world. Rapid industrialization, energy production and increased trade, which began with the industrial revolution, have promoted economic growth while causing irreversible damage to nature, ecosystems and agricultural activities. In particular, the overuse of fossil fuels has led to a rapid increase in greenhouse gases released into the atmosphere and deepened environmental degradation. This has led to widespread environmental problems such as global warming, climate change, air and water pollution. These problems have become a threat to the ecosystem (Agenor, 2004; Madaleno & Nogueira, 2023).

The industrial revolution led to the growth in economic activity, mass production and a rapid increase in energy consumption. In particular, the widespread use of fossil fuels such as coal, oil and natural gas has become the engine of economic growth, but has also been the main source of environmental problems. The combustion of fossil fuels releases carbon dioxide ($CO_2$) into the atmosphere. $CO_2$ emissions have triggered global warming by causing the greenhouse effect, leading to climate change and environmental disasters. The increase in energy demand and industrial activities in rapidly industrializing countries such as Türkiye has led to the degradation of ecosystems and the decrease in biodiversity (Bilen et al., 2008). In this process, especially ecosystem has suffered great damage. Industry and energy production have led to the destruction of natural habitats, deforestation and rapid depletion of natural resources. Ecosystems and biodiversity have come under great pressure due to these economic activities. Air, water and soil pollution in industrialized regions in Türkiye has directly affected agricultural areas and ecosystems. This pollution both threatens the sustainability of ecosystems and reduces productivity in agricultural production (Richmond & Kaufmann, 2006).

Agricultural activities are one of the areas most affected by economic growth. Rapid urbanization and the expansion of industrial areas have led to shrinking agricultural areas and declining productivity. Misuse of agricultural land, especially overconsumption of water resources and intensive cultivation of land threaten the sustainability of agriculture. In developing countries such as Türkiye, rapidly increasing energy demand and industrial production put serious pressure on agricultural land. Agricultural production has declined in many regions where access to water resources has become more difficult and soil fertility has declined. As a result, food security risks have increased. Moreover, climate change triggered by the intensive use of fossil fuels has disrupted rainfall patterns and created uncertainty in agricultural activities (Ulucak & Erdem, 2012).

The relationship between economic growth and environmental degradation is not limited to agriculture and energy consumption. Excessive consumption of natural resources has led to permanent damage to the environment, loss of biodiversity and disruption of ecosystem balance. Fossil fuels used to meet energy needs have led to the release of large amounts of greenhouse gases into the atmosphere. This has accelerated global warming and caused irreversible impacts on both nature and human life (Ansuategi & Escapa, 2002).

Global warming, climate change and environmental degradation lead to a deterioration of the natural balance around the world. This process threatens the ecosystems and leads to the degradation of natural systems. Climate change is especially putting pressure on the agricultural sector. Changes in rainfall patterns, droughts and extreme temperatures adversely affect agricultural production and put food security at risk. Moreover, environmental pollution and depletion of natural resources weaken the self-renewal capacity of ecosystems (Jacobs et al., 2014). Various attempts have been made internationally to solve these environmental problems. The 1972 Stockholm Conference and international initiatives such as the Kyoto Protocol are important steps to ensure environmental sustainability. However, the impact of these initiatives has been limited. Generally, economic growth-oriented policies have overshadowed environmental sustainability. Developing countries in particular have failed to adequately implement environmental protection policies while achieving their economic growth targets (Nordhaus, 2010). Türkiye is one of the countries involved in this process. Although Türkiye joined the Kyoto Protocol in 2009 and committed to reduce greenhouse gas emissions, there is still a large imbalance between growth targets and environmental protection (Turhan et al., 2016).

As a result, the environmental impacts of economic growth are a major threat to the sustainability of ecosystems, agricultural production and nature. In developing countries like Türkiye, it is becoming increasingly difficult to strike a balance between economic growth and environmental protection. Rapidly increasing energy demand and industrial activities threaten ecosystem and the sustainability of natural resources. In this process, it has become imperative to implement environmental protection policies more effectively, reduce energy consumption and turn to renewable energy sources. For economic growth to be sustainable, natural resources must be managed effectively and the environment must be protected.

In this regard, the impacts of economic growth on ecosystems require a more detailed analysis. Therefore, this study investigates the impacts of economic growth on the ecosystem and its components. The Ecosystem Vitality index, a sub-dimension of the Environmental Performance Index, is used in the study. The ecosystem vitality index includes key indicators such as biodiversity and habitat, ecosystem loss, fisheries, acid rain, agriculture and water resources. These indicators represent different dimensions of environmental degradation. The aim of the study is to identify how economic activities damage nature through these different ecosystem components and to discuss what steps should be taken to mitigate these damages.

This study will contribute to the literature in at least three ways. (1) To the best of our knowledge, there is no empirical study investigating the impact of economic growth on different ecosystem components in Türkiye. This study analyzes the impacts of economic growth on ecosystem to fill this gap in the literature. (2) In order to determine the impacts of economic growth on different components of the ecosystem, all components are analyzed separately. For this purpose, seven different econometric models were constructed. This provided a clearer analysis of the potential impact of the economic model on each ecosystem component. (3) We used the most recent dataset available, thus providing a real-time and up-to-date perspective.

The rest of the paper is organized as follows: Section 2 presents the literature review, Section 3 presents the data and methodology, Section 4 presents the empirical findings, Section 5 presents policy implications and finally Section 6 presents conclusions.

## LITERATURE REVIEW

Human activities, particularly industrialization and urbanization, create serious pressures on natural resources and lead to environmental pollution. Economic growth is regarded as one of the most important sources of direct environmental degradation. Many studies have shown that economic growth damages the environment through factors such as increased energy demand, resource use and waste generation. However, the protection of the environment and nature is of great importance for both the ecosystem sustainability and human health. Therefore, the relationship between economic activities and their environmental impacts has been extensively analyzed in the literature. However, the relationship between economic growth and environmental pollution is a complex issue with different results in the literature. Among these studies, Grossman and Krueger (1993) have been pioneers in this field with their studies suggesting that economic growth may initially increase environmental pollution, but may decrease pollution after a certain level of income. Lopez (1994) also focused on the environmental impacts of economic growth and examined how growth and trade liberalization affect the environment. In more recent studies, Al-mulali (2012) emphasized that energy consumption and growth have a positive impact on $CO_2$ emissions and that foreign trade and FDI strengthen this relationship. Moreover, Mahmood et al. (2019) emphasize the emissions-enhancing impact of growth and trade liberalization in Tunisia. However, some studies have found that the impacts of growth on environmental degradation are more limited.

The impacts of energy consumption and economic growth on $CO_2$ emissions stand out. Jalil and Mahmud (2009) examined the impacts of economic growth, energy consumption and trade on $CO_2$ emissions in China. The results show that economic growth increases $CO_2$ emissions. Similarly, Jayanthakumaran et al. (2012) found that growth and energy consumption increase $CO_2$ emissions in India and China. Al-mulali and Sheau-Ting (2014) pointed out the positive impact of energy consumption and foreign trade on $CO_2$ emissions. Shahbaz et al. (2013) find that growth and energy consumption have a positive impact on $CO_2$ emissions in Indonesia, while financial development and trade have a negative impact. Moreover, other studies such as Farhani et al. (2014) and Bozkurt and Okumus (2017) also conclude that energy consumption increases emissions.

In addition these studies examining the impacts of economic growth on the environment, the literature on the environmental impacts of trade is also quite extensive. Studies on the environmental impacts of trade focus especially on the role of free trade on environmental pollution. Copeland and Taylor (1994) showed that trade can increase pollution with a two-country static general equilibrium model. More recently, Antweiler et al. (2001) argued that trade can both increase and decrease pollution. He argued that this relationship should be addressed through scale, technical and composition effects. Moreover, Rock (1996) emphasized that the impacts of trade on the environment differ in developed and developing countries and argued that open policies in particular can increase pollution intensity. Studies such as Baek et al. (2009) have shown that trade intensity is positively associated with environmental degradation in developing countries. Moreover, an increase in income level strengthens this impact. Kukla-Gryz (2009) evaluated the impacts of trade and per capita income on air pollution and found that trade increases pollution especially in developing countries.

In recent years, ecological footprint has become a more frequently used measure among environmental indicators in the literature. Gao and Tian (2016) examined the impacts of trade on ecological footprint in China. They found that China is an importer of ecological footprint. This finding suggests that large economies such as China should reconsider their trade policies in terms of environmental degradation. Usman et al. (2020) find that trade openness has a negative impact on ecological footprint in Africa, Asia and the Americas. They also find that there is a unidirectional causality from trade openness to ecological footprint. Rehman et al. (2021) examined the impacts of trade and energy consumption on ecological footprint in Pakistan and found that trade increases environmental degradation in both the short- and long-run. Yilanci et al. (2022) examined the relationship between trade openness and ecological footprint in G7 countries and found that trade openness has complex impacts on ecological footprint.

The complexity of the impacts of economic growth on the environment stems from the multi-dynamic nature of the growth process. This leads to the emergence of various approaches in the literature. Among these approaches, the Environmental Kuznets Curve (EKC) hypothesis argues that there is an inverted-U relationship between economic growth and environmental pollution. The EKC suggests that economic growth increases environmental pollution at low income levels, but after a certain income level, growth starts to improve environmental quality. Among the studies investigating the EKC, Kasman and Duman (2015) find that the EKE hypothesis is valid in EU countries. They showed that economic growth initially increases $CO_2$ emissions, but this effect reverses as income increases. Farhani et al. (2014) obtained similar findings for Tunisia and emphasized that trade liberalization and energy consumption also increase emissions. Moreover, Dogan et al. (2017) find that the ECA hypothesis is not valid in OECD countries. The study finds that energy consumption and tourism increase emissions, but trade openness decreases emissions.

Various methods are used to evaluate the impact of different factors on the environment. One of the most prominent of these is the Environmental Performance Index (EPI). The EPI provides a comprehensive tool for measuring and analyzing environmental sustainability performance. These studies play an important guiding role in the process of developing countries' environmental policies. For example, in a study conducted by Alptekin (2015), Türkiye's sustainability performance was compared with the European Union member states. The study shows that Sweden has the highest performance, while Türkiye ranks 20th. These findings reveal that Türkiye is far behind the European Union average in terms of environmental performance and that more efforts should be made in this area. Similarly, Karaman (2018) evaluated Türkiye's environmental performance in line with the European Union membership target. The results revealed that Türkiye's environmental policies are quite weak compared to European standards. Another study detailing Türkiye's poor environmental performance was conducted by Bek (2019). In this study, the environmental performance of Switzerland and Türkiye is compared and the reasons for Türkiye's low ranking are investigated. Pimonenko et al. (2018) analyzed the methodology of the EPI and argued that this index is an important tool for assessing the environmental, social and economic status of countries. In addition, it was revealed that countries with high performance in the EPI are also successful in the Sustainable Development Goals and the Social Progress Index. Another study by Ozkan and Ozcan (2018) evaluated the environmental performance of OECD countries. In this study, Türkiye's environmental performance was found to be in an effective position. However, when the results are compared with previous studies, Türkiye's environmental performance ranking remains relatively low. Pinar (2022) examined the sensitivity of environmental performance indicators to the subjective weights assigned to them using EPI data. The study emphasizes that sensitivity analysis of environmental performance is critical for reliability and transparency. Iskenderoglu et al. (2023) also examined the environmental impacts of economic growth, foreign direct investment and renewable energy consumption using the Environmental Performance Index. The results of this study show that FDI plays a positive role in reducing environmental degradation, while economic growth has a negative impact. This shows that the impacts of economic factors on the environment are complex and multidimensional.

All these studies show that the EPI is a useful tool for comparing countries' environmental performance on a global scale. These analyses of various countries and regions emphasize the importance of the methodological approaches used to assess countries' environmental performance. While studies such as Alptekin (2015) and Karaman (2018) reveal Türkiye's low environmental performance, studies such as Pimonenko et al. (2018) and Pinar (2022) highlight the methodological sensitivities of the EPI and the importance of sensitivity analysis. Studies such as Ozkan and Ozcan (2018) and Iskenderoglu et al. (2023) reveal the complexity in this area by analyzing the relationship between environmental performance and economic and financial variables.

## MATERIAL AND METHODS
### Model Specification and Data

The empirical analysis of this study analyzes the impacts of economic growth on ecosystem in Türkiye. The study uses annual data for the period 1995-2021. This period was chosen considering the availability of the dataset and its suitability for the analysis. Descriptive statistics of the data used in the study are presented in Table 1.

Turkish Journal of Agricultural and Natural Sciences 12 (2): 397–411, 2025

**Table 1.** Descriptions and sources of variables

| Variable | Notation | Description | Obs. | Mean | SD | Min. | Max. | Source |
|---|---|---|---|---|---|---|---|---|
| Biodiversity & habitat | BDH | Index | 27 | 1.042 | 0.006 | 1.025 | 1.050 | EPI |
| Ecosystem services | ECS | Index | 27 | 1.492 | 0.091 | 1.343 | 1.624 | EPI |
| Fisheries | FSH | Index | 27 | 1.157 | 0.163 | 0.841 | 1.374 | EPI |
| Acid rain | ACD | Index | 27 | 1.808 | 0.087 | 1.680 | 1.974 | EPI |
| Agriculture | AGR | Index | 27 | 1.564 | 0.024 | 1.505 | 1.605 | EPI |
| Water resources | WRS | Index | 27 | 1.484 | 0.002 | 1.475 | 1.485 | EPI |
| Ecosystem vitality | ECO | Index | 27 | 1.388 | 0.027 | 1.338 | 1.436 | EPI |
| GDP per capita | GDP | Constant 2015 US$ | 27 | 3.925 | 0.115 | 3.756 | 4.129 | WB |
| Energy intensity | ENG | Level of primary energy | 27 | 0.466 | 0.043 | 0.394 | 0.537 | WB |
| Population density | POP | People per sq. km of land | 27 | 1.966 | 0.045 | 1.887 | 2.039 | WB |
| Trade | TRD | Trade (% of GDP) | 27 | 1.703 | 0.059 | 1.576 | 1.852 | WB |

Note: (1) WB and EPI indicate World Bank-World Development Indicators, and Yale University-Environmental Performance Index, respectively. (2) The abbreviations N, SD, SE, Min and Max denote the number of observations, standard deviation, standard error, minimum and maximum values, respectively. (3) The variables used in the study are logarithmized.

Ecosystem vitality index was used as the dependent variable in the study. Ecosystem vitality includes factors such as sustainability of natural resources and conservation of biodiversity and is an important indicator for monitoring environmental degradation. The ecosystem vitality index is a sub-dimension of the Environmental Performance Index (EPI) developed by Wolf et al. (2022) of Yale University. The EPI is a comprehensive index that aims to measure a country's environmental sustainability performance. It is widely used by governments, policy makers and researchers worldwide. The EPI consists of three main dimensions: Climate Change, Environmental Health and Ecosystem Vitality. Each dimension includes a set of components to measure different areas of environmental performance. The purpose of the EPI is to provide a benchmarking tool for countries to develop environmentally sound policies and improve their implementation. The components of the ecosystem vitality index are given in Table 2.

**Table 2.** Components of the ecosystem vitality index

| Dimension | Weight | Indicator | Notation | Weight |
|---|---|---|---|---|
| Biodiversity & Habitat (BDH) | 43.00% | Terrestrial Biome Protection (national) | TBN | 22.2% |
| | | Terrestrial Biome Protection (global) | TBG | 22.2% |
| | | Marine Protected Areas | MPA | 22.2% |
| | | Protected Areas Rep. Index | PAR | 14% |
| | | Species Habitat Index | SHI | 8.3% |
| | | Species Protection Index | SPI | 8.3% |
| | | Biodiversity Habitat Index | BHV | 3% |
| Ecosystem Services (ECS) | 19.00% | Tree Cover Loss | TCL | 75% |
| | | Grassland Loss | GRL | 12.5% |
| | | Wetland Loss | WTL | 12.5% |
| Fisheries (FSH) | 11.90% | Fish Stock Status | FSS | 36% |
| | | Marine Trophic Index | RMS | 36% |
| | | Fish Caught by Trawling | FTD | 28% |
| Acid Rain (ACD) | 9.50% | SO₂ Growth Rate | SDA | 50% |
| | | NOₓ Growth Rate | NXA | 50% |
| Agriculture (AGR) | 9.50% | Sustainable Nitrogen Mgmt. Index | SNM | 50% |
| | | Sustainable Pesticide Use | SPU | 50% |
| Water Resources (WRS) | 7.10% | Wastewater Treatment | WWVT | 100% |

Source: Wolf et al. (2022)

Table 2 shows that the ecosystem vitality index is composed of several components that assess the sustainability of nature conservation and ecosystem services. The biodiversity and habitat dimension (43%) constitutes the largest part of this index. Here, indicators such as the protection of terrestrial biomes at both national and global level, marine protected areas and the representativeness of protected areas are considered. The ecosystem services dimension (19%) tracks the loss of tree cover, grasslands and wetlands, while the fisheries dimension (11.9%) measures the state of fish stocks and the impacts of trawling on the ecosystem. It also includes environmental pressures such as acid rain (9.5%), sustainability of agricultural practices (9.5%) and management of water resources (7.1%). These components comprehensively analyze ecosystem vitality and environmental performance.

The main objective of this study is to analyze the impacts of economic growth on ecosystem. For this purpose, the Ecosystem Vitality Index, a sub-dimension of the Environmental Performance Index, was used in the study. Not only the overall ecosystem vitality but also the impacts on the sub-components of ecosystem vitality are analyzed. Seven models are created to evaluate in detail the impact of economic growth on different dimensions of ecosystem vitality. Models 1 to 6 explore the impacts of economic growth on the sub-components of ecosystem vitality (biodiversity and habitat, ecosystem services, fisheries, acid rain, agriculture and water resources). Model 7 aims to analyze the impacts on total ecosystem vitality.

While constructing the models used in the study, the shortcomings of empirical studies in the relevant literature were considered. Control variables were also included in the analysis to provide a more comprehensive analysis of the factors on ecosystem vitality. Based on the literature review, GDP per capita (Agboola et al., 2021; Tabash et al., 2024), energy intensity (Jalil and Mahmud, 2009; Al-muali, 2012; Jayanthakumaran et al., 2012), population density (Ghanem, 2018; Van Dao and Van, 2020), and trade (Sheau-Ting, 2014; Mahmood et al., 2019) were identified as control variables. These models are shown as follows.

$$BDH_t = \beta_0 + \beta_1 GDP_t + \beta_2 ENG_t + \beta_3 POP_t + \beta_4 TRD_t + \epsilon_t \quad (1)$$
$$ECS_t = \beta_0 + \beta_1 GDP_t + \beta_2 ENG_t + \beta_3 POP_t + \beta_4 TRD_t + \epsilon_t \quad (2)$$
$$FSH_t = \beta_0 + \beta_1 GDP_t + \beta_2 ENG_t + \beta_3 POP_t + \beta_4 TRD_t + \epsilon_t \quad (3)$$
$$ACD_t = \beta_0 + \beta_1 GDP_t + \beta_2 ENG_t + \beta_3 POP_t + \beta_4 TRD_t + \epsilon_t \quad (4)$$
$$AGR_t = \beta_0 + \beta_1 GDP_t + \beta_2 ENG_t + \beta_3 POP_t + \beta_4 TRD_t + \epsilon_t \quad (5)$$
$$WRS_t = \beta_0 + \beta_1 GDP_t + \beta_2 ENG_t + \beta_3 POP_t + \beta_4 TRD_t + \epsilon_t \quad (6)$$
$$ECO_t = \beta_0 + \beta_1 GDP_t + \beta_2 ENG_t + \beta_3 POP_t + \beta_4 TRD_t + \epsilon_t \quad (7)$$

In these equations, $\beta_0$ epresents the constant term, while the coefficients $\beta_1$ to $\beta_4$ measure the impact of each independent variable on exports. $\epsilon_t$ is the error term with zero mean and constant variance, where t is the time period.

### Unit Root Analysis

In this study, two common unit root tests are applied to determine the stationarity properties of time series data: Augmented Dickey-Fuller (ADF) Test and Phillips-Perron (PP) Test. Both tests determine whether the series are stationary or not, thus allowing the existence of long-run relationships and the selection of appropriate econometric methods.

**ADF Unit Root Test**: The ADF test developed by Dickey and Fuller (1979) is one of the most widely used methods to test whether a time series is stationary. The ADF test extends the Dickey-Fuller (DF) test to include lagged difference terms in the model to eliminate the autocorrelation problem. In this way, it provides more reliable results by ensuring the independence assumption. The general equation of the ADF test is as follows:

$$\Delta Y_t = \alpha + \beta_t + \gamma Y_{t-1} + \sum_{i=1}^{p} \delta_i \Delta Y_{t-i} + \epsilon_t \quad (8)$$

In Equation 8, $Y_t$ is the time series under test, $\Delta Y_t$ is the first difference of the series, $\alpha$ is the constant term, $\beta_t$ is the time trend, $\gamma Y_{t-1}$ is the coefficient testing for the presence of a unit root and $\epsilon_t$ is the error term. In the ADF test, a coefficient γ equal to zero indicates that there is a unit root in the series and the series is non-stationary.

**PP Unit Root Test**: The PP test developed by Phillips and Perron (1988) is an alternative approach to the ADF test. The PP test considers the presence of autocorrelation and heteroskedasticity in determining whether the series are stationary. Similar to the ADF test, the PP test tests the stationarity of the series. However, instead of adding lag terms to the model, it addresses the problems of autocorrelation and heteroskedasticity by making asymptotic corrections to the error terms (Perron, 1988). The general equation of the PP test is as follows:

$$\Delta Y_t = \alpha + \beta_t + \gamma Y_{t-1} + \epsilon_t \quad (9)$$

In Equation 9, $Y_t$ is the level of the series, $\Delta Y_t$ is the first difference, $\alpha$ is the constant term, $\beta_t$ is the time trend, $\gamma Y_{t-1}$ is the lagged value and $\epsilon_t$ is the error term. The null hypothesis of the PP test is the existence of a unit root in the series. A coefficient of γ equal to zero indicates that the series contains a unit root and is non-stationary.

The main objective of both tests is to test whether there is a unit root in the series. In this context, the null hypothesis ($H_0$) states that the series has a unit root, i.e. is non-stationary, while the alternative hypothesis ($H_1$) states that the series is stationary. The lag length is usually determined by criteria such as Akaike Information Criterion (AIC) or Schwarz Information Criterion (SIC). Since the PP test solves the problems of autocorrelation and heteroskedasticity with asymptotic methods, it is more flexible and can produce more robust results for large data sets. It also provides reliable results by correcting the error terms without changing the parameter estimates.

## ARDL Cointegration Test

There are various methods to test cointegration between series. Among these methods, the tests developed by Engle and Granger (1987), Johansen (1988) and Johansen and Juselius (1990) stand out. However, these methods have some limitations. The ARDL model developed by Pesaran et al. (2001) is a method developed to test the long run relationship between variables. It stands out by eliminating the limitations of the mentioned methods. In particular, the fact that it does not require the variables to be stationary at the same level makes the ARDL method a more flexible option. This method has the capacity to analyze stationary variables at both I(0) and I(1) levels. However, I(2) variables should not be included in the model. This method provides a significant advantage in econometric analysis by providing reliable results even in data sets with a limited number of observations. By including lags of dependent and independent variables in the model, short run dynamics and long run relationships can be analyzed together. With these features, the ARDL bounds test is an effective method for determining cointegration relationships. The ARDL model is expressed in the following general form:

$$\Delta Y_t = \alpha + \sum_{i=1}^{p} \beta_i \Delta Y_{t-i} + \sum_{j=0}^{q} \theta_j \Delta X_{t-j} + \phi Y_{t-1} + \psi X_{t-1} + \epsilon_t \qquad (10)$$

In Equation 10, $Y_t$ is the dependent variable, $X_t$ is the independent variables, $\Delta Y_t$ and $\Delta X_t$ are the first differences of the series, $\beta_i$ and $\theta_j$ are the coefficients of short-run dynamics. $\phi Y_{t-1}$ and $\psi X_{t-1}$ are the coefficients of long-run relationships and $\epsilon_t$ is the error term. The ARDL cointegration bounds test tests whether the long-run coefficients are zero. This method examines whether there is a long-run relationship between variables and estimates the long-run coefficients under appropriate conditions.

The ARDL cointegration test tests whether there is a long-run relationship (cointegration) between variables. The null hypothesis ($H_0$), states that there is no long-run relationship, while the alternative hypothesis ($H_1$) states that there is a long-run relationship. The test is performed using the F-statistic. The F-statistic is compared with certain critical values. If the F-statistic is less than the lower bound, it is concluded that there is no cointegration; if it is greater than the upper bound, cointegration is accepted. When the F-statistic falls between these two limits, the result is ambiguous.

## ARDL Coefficient Estimates

After determining the existence of cointegration, it is possible to analyze the long-run relationship between variables using the ARDL model. Long-run coefficients are calculated through the coefficients obtained from the ARDL model. In the ARDL model, the long-run impacts of the independent variables are calculated by the $\psi/\phi$ ratio. The long-run form of the model is expressed as follows:

$$Y_t = \alpha + \sum_{j=0}^{q} \theta_j X_{t-j} + \epsilon_t \qquad (11)$$

In Equation 11, $Y_t$ is the dependent variable, $X_{t-j}$ s the lagged values of the independent variables and $\epsilon_t$ is the error term. The ARDL model allows the estimation of both short-run dynamics and long-run relationships. Long-run coefficients determine the strength and direction of long-run relationships in the model.

Long-run coefficient estimation is done by normalizing the coefficients of lagged independent variables in the ARDL model by the error correction term. In this way, the effect of a one-unit increase in the independent variables on the dependent variable in the long-run is determined. The ARDL model provides a dynamic structure by simultaneously analyzing the long-run equilibrium relationship and short-run deviations.

## EMPIRICAL FINDINGS

ADF unit root test and PP unit root test were applied to determine whether the variables used in the study contain unit roots. Moreover, ARDL bounds test was used to analyze the long-run cointegration relationship between the variables. To estimate the long-run coefficients, the ARDL model was used to estimate the long-run impacts. The results obtained with these methods are given as follows.

**Table 3.** Unit root test results

| Variable | | ADF unit root test | | PP unit root test | |
|---|---|---|---|---|---|
| | | t-statistic (level) | t-statistic (first difference) | t-statistic (level) | t-statistic (first difference) |
| BDH | Constant | -1.133 | -3.774*** | -2.512 | -3.775*** |
| ECS | | -0.325 | -4.133*** | -0.325 | -4.109*** |
| FSH | | -1.094 | -4.412*** | -1.043 | -3.984*** |
| ACD | | -2.397 | -3.476*** | -1.834 | -2.279*** |
| AGR | | -2.109 | -6.063*** | -1.847 | -5.156*** |
| WRS | | -1.460 | -7.163*** | -1.444 | -7.208*** |
| ECO | | -2.199 | -3.051** | -1.099 | -2.617*** |
| GDP | | -0.188 | -4.456*** | -0.188 | -4.456*** |
| ENG | | -0.819 | -4.951*** | -0.629 | -5.422*** |
| POP | | -2.313*** | -4.024*** | -1.641*** | -4.038*** |
| TRD | | -1.234 | -4.865*** | -0.934 | -5.586*** |
| BDH | Constant and Trend | -1.619 | -4.434*** | -1.511 | -4.434*** |
| ECS | | -2.342 | -4.302*** | -2.332 | -4.272*** |
| FSH | | -3.163 | -4.307*** | -2.403 | -3.849** |
| ACD | | -2.324 | -4.025*** | -1.509 | -2.371*** |
| AGR | | -3.819 | -5.916*** | -3.147 | -5.188*** |
| WRS | | -1.865 | -7.462*** | -1.865 | -7.177*** |
| ECO | | -3.069 | -3.854*** | -1.658 | -2.753*** |
| GDP | | -2.716 | -4.444*** | -2.227 | -4.435*** |
| ENG | | -2.891 | -4.839*** | -3.007 | -5.246*** |
| POP | | -2.145*** | -4.241*** | -1.663*** | -5.098*** |
| TRD | | -2.795 | -6.761*** | -2.312 | -5.947*** |

Note: The superscripts ***, **, and * denote the significance at a 1%, 5%, and 10% level, respectively.

Table 3 shows that all variables except POP in the constant model contain unit root at level. However, these variables become stationary in their first differences. This shows that all variables except POP are I(I). The POP variable is determined as I(0) since it is stationary at level. Similar findings were obtained in the analysis based on the constant and trend model. Variables other than POP is not stationary at level. However, variables become stationary when first differences are taken.

As a result, POP variable was found to be stationary at level. All other variables become stationary when first differences are taken. This means that POP is I(0) and the other variables are I(I). Since these findings show that the series are stationary at different levels, it allows us to use the ARDL (Autoregressive Distributed Lag) method. Because the ARDL model is suitable for working with both level stationary variables and stationary variables at first difference. In addition, it is an ideal method for analyzing series with mixed unit root properties.

**Table 4.** ARDL bound test results

| Model | Optimal lag length | F-statistics | Critical values %5 | | Critical values %1 | |
|---|---|---|---|---|---|---|
| | | | I(0) | I(I) | I(0) | I(I) |
| Model 1: F(BDH \| GDP, ENG, POP, TRD) | (2, 2, 1, 1, 1) | 4.192** | 2.56 | 3.49 | 3.29 | 4.37 |
| Model 2: F(ECS \| GDP, ENG, POP, TRD) | (1, 0, 0, 1, 0) | 5.254*** | 2.56 | 3.49 | 3.29 | 4.37 |
| Model 3: F(FSH \| GDP, ENG, POP, TRD) | (2, 0, 1, 0, 0) | 1.282* | 2.56 | 3.49 | 3.29 | 4.37 |
| Model 4: F(ACD \| GDP, ENG, POP, TRD) | (2, 0, 1, 0, 0) | 3.979** | 2.56 | 3.49 | 3.29 | 4.37 |
| Model 5: F(AGR \| GDP, ENG, POP, TRD) | (2, 1, 0, 1, 2) | 8.342*** | 2.56 | 3.49 | 3.29 | 4.37 |
| Model 6: F(WRS \| GDP, ENG, POP, TRD) | (2, 2, 0, 1, 1) | 4.282** | 2.56 | 3.49 | 3.29 | 4.37 |
| Model 7: F(ECO \| GDP, ENG, POP, TRD) | (2, 2, 2, 1, 2) | 4.558*** | 2.56 | 3.49 | 3.29 | 4.37 |

Note: (1) The superscripts ***, **, and * denote the significance at a 1%, 5%, and 10% level, respectively. (2) The optimal lag lengths are calculated automatically using information criteria.

The results in Table 4 show that there is a significant cointegration relationship at the 5% level in Model 1, Model 4 and Model 6. In these models, F-statistics are above the critical values at the 5% level, indicating that there is a long-run relationship between the variables. Moreover, there is a significant cointegration relationship at the 1% level in Model 2, Model 5 and Model 7. In these models, the F-statistics are above the 1% critical values, indicating a stronger cointegration relationship. In Model 3, the cointegration relationship is found at the 10% significance level and the long-run relationship in this model is weaker than in the other models. These results indicate that there is a long-run cointegration relationship between variables in most of the models analyzed with the ARDL bounds test. These findings confirm that the ARDL model is appropriate for analyzing the dynamics between economic growth and ecosystem vitality.

The results of the diagnostic tests conducted to assess the validity and reliability of the model are presented in Table 5. The results of the diagnostic tests show that there are no statistical problems in the models. Moreover, CUSUM and CUSUM Square tests, which assess the stability of long-run estimates, confirm that the model parameters are stable over the sample period. Moreover, the coefficient ECT(-1) in the models is called the error correction term and represents the speed of return to long-run equilibrium. As expected, the coefficient of ECT(-1) is negative and statistically significant in all models. A negative sign indicates that short-run imbalances are corrected over time and the system returns to long-run equilibrium.

**Table 5.** Short-run and long-run estimation.

| Variable | Short-run coefficient | | | | | | |
|---|---|---|---|---|---|---|---|
| | Model 1 | Model 2 | Model 3 | Model 4 | Model 5 | Model 6 | Model 7 |
| | $BDH_{(M-D)}$ | $ECS_{(M-D)}$ | $FSH_{(M-D)}$ | $ACD_{(M-D)}$ | $AGR_{(M-D)}$ | $WRS_{(M-D)}$ | $ECO_{(M-D)}$ |
| GDP | -0.253* | -0.101* | -0.533** | -0.526** | 0.521*** | 0.362 | 0.044*** |
| ENG | 0.012 | -0.502*** | 1.659 | -0.776*** | -0.073*** | -0.744** | -0.528*** |
| POP | -0.677*** | -5.044*** | -1.161*** | 1.236 | 0.557** | -1.739*** | 1.339*** |
| TRD | -0.022** | -0.077*** | -0.213** | -0.006** | 0.106 | -1.562 | -0.103*** |
| ECT(-1) | -0.325*** | -0.358*** | -0.176*** | -0.449*** | -0.312*** | -0.527*** | -0.349*** |
| Variable | Long-run coefficient | | | | | | |
| | Model 1 | Model 2 | Model 3 | Model 4 | Model 5 | Model 6 | Model 7 |
| | $BDH_{(M-D)}$ | $ECS_{(M-D)}$ | $FSH_{(M-D)}$ | $ACD_{(M-D)}$ | $AGR_{(M-D)}$ | $WRS_{(M-D)}$ | $ECO_{(M-D)}$ |
| GDP | -0.116*** | -0.591** | -1.362** | -2.144*** | 0.074*** | 0.672*** | -0.668*** |
| ENG | -0.081** | -1.378*** | -0.505 | -3.960*** | -0.183*** | -1.442*** | -1.786*** |
| POP | -0.336** | -1.857*** | -0.668** | -1.481** | -0.392** | -1.870*** | -0.047* |
| TRD | -0.083*** | -0.225** | -0.319** | -2.203*** | -0.010 | -1.800 | -0.221** |
| C | 1.021*** | 8.607* | 7.528*** | -0.464*** | 0.457*** | 2.245* | 5.286*** |
| Diagnostic test | P value | P value | P value | P value | P value | P value | P value |
| Serial correlation | 0.47 | 0.34 | 0.27 | 0.55 | 0.29 | 0.51 | 0.15 |
| Heteroskedasticity | 0.57 | 0.32 | 0.22 | 0.65 | 0.78 | 0.24 | 0.86 |
| Normality | 0.79 | 0.75 | 0.47 | 0.37 | 0.81 | 0.67 | 0.11 |
| Functional form | 0.92 | 0.57 | 0.72 | 0.57 | 0.63 | 0.14 | 0.64 |
| CUSUM | Stable | Stable | Stable | Stable | Stable | Stable | Stable |
| CUSUMSQ | Stable | Stable | Stable | Stable | Stable | Stable | Stable |

Note: The superscripts ***, **, and * denote the significance at a 1%, 5%, and 10% level, respectively.

The results show that the impact of GDP on Model 1, Model 2, Model 3, Model 4 and Model 7 is negative and statistically significant. In these models, a 1% increase in GDP reduces biodiversity and habitat, ecosystem services, fisheries, acid rain and total ecosystem vitality by 0.101-2.144%. However, the impact of GDP on Model 5 and Model 6 is found to be positive and statistically significant. In these models, a 1% increase in GDP increases agricultural and water resources by 0.074-0.672%. This shows that economic growth increases pressures on environmental sustainability and ecosystems.

The negative impact of GDP on biodiversity and habitat shows that economic growth increases consumption of natural resources, leading to habitat loss. Economic growth in developing countries is often based on industrial, mining and agricultural expansion, which leads to the destruction of natural habitats. This process leads to the degradation of natural ecosystems and a decline in biodiversity. The negative impact of GDP on ecosystem services shows that economic growth can damage environmental services. For example, during periods of growth, activities such as agriculture, industry and energy production can negatively affect soil, water and air quality. Economic growth can lead to the degradation of services provided by ecosystems (water purification, air purification, etc.), creating long-run environmental and social costs. The negative impact of GDP on fisheries shows that economic growth also puts marine ecosystems under pressure. Economic growth, together with trade and industrialization, can lead to overexploitation of fish stocks. This threatens the sustainability of the fisheries sector by reducing marine biodiversity. The negative impact of GDP on acid rain shows that economic growth increases the use of fossil fuels, which in turn increases harmful emissions to the environment. Growth based on fossil fuels releases gases such as sulphur dioxide and nitrogen oxides into the atmosphere, causing acid rain. Acid rain has destructive impacts on forests, lakes and buildings. The negative impact of GDP on total ecosystem vitality shows that economic growth has a negative impact on the overall health of ecosystems. Economic growth can undermine ecosystem vitality by leading to overuse of natural resources through agriculture, industry and urban expansion. In the long-run, this can lead to ecosystem degradation and environmental sustainability problems.

Model 5 (agriculture) and Model 6 (water resources) are the models where a positive impact of GDP is observed. This suggests that economic growth may have some positive impacts in these areas. The positive impact of GDP on agriculture suggests that economic growth can increase agricultural production. Economic growth can lead to positive outcomes such as strengthening agricultural infrastructure and investing in technology and mechanization. In particular, agricultural productivity growth can provide significant benefits in terms of food security and economic development. However, if growth is not supported by sustainable agricultural policies, environmental costs may arise. The positive impact of GDP on water resources suggests that economic growth can contribute to the development of water management infrastructure. Investments in water treatment facilities can increase the environmental benefits of economic growth. However, this positive impact needs to be supported by policies for sustainable water use and protection of water resources. Otherwise, increased industrial and agricultural activities may put pressure on water resources.

Additionally, the study uses three explanatory variables: energy intensity, population and trade. First, energy intensity has a negative impact in almost all models and is statistically significant in most models. This clearly shows the negative impact of energy intensity on environmental indicators. As energy intensity is associated with fossil fuel consumption, it can increase emissions that are particularly harmful for the environment, leading to habitat destruction, reduced ecosystem services and increased acid rain. Moreover, while the negative impact of energy intensity in agriculture is less pronounced, energy use can cause environmental pressures, especially when agriculture is not sustainable. Second, population density has a negative impact in most models and is statistically significant. Population growth implies greater use of environmental resources and a reduction in living space. As the population increases, energy consumption, waste production and demand for natural resources also increase. This leads to environmental destruction. The negative impact of population growth on water resources is noteworthy. As the population increases, the pressure on water resources increases and sustainable water management becomes more difficult. Third and finally, trade has a negative impact in most models and is statistically significant. This reveals the negative impacts of trade on the environment. Increased trade volume increases the use of fossil fuels, especially by increasing transportation activities. Therefore, emissions harmful to the environment also increase. Moreover, the increase in trade volume may also contribute to unsustainable production processes. The negative impact of trade on acid rain points to damage to marine ecosystems and the environment in general.

The results show that economic growth has a positive impact on agriculture and water resources. This finding suggests that economic growth can stimulate the adoption of technological innovations and modern farming methods to increase agricultural production and productivity. Investments in the agricultural sector during economic growth can strengthen agricultural infrastructure, increase mechanization and introduce efficient irrigation systems. This can have positive impacts on food security, agricultural production capacity and rural development. Moreover, investments in water resources can improve water productivity in both urban and rural areas. However, a negative impact of GDP on biodiversity and habitat, ecosystem services, fisheries, acid rain and total ecosystem vitality was observed. These findings suggest that economic growth threatens environmental sustainability in the long-run, particularly through overconsumption of natural resources and environmental pressures. Energy intensity, population and trade generally have a negative impact on the environmental indicators in the models. The main reasons for this are excessive energy consumption, reliance on fossil sources of energy and neglect of environmental sustainability. In particular, the environmental costs of fossil fuel use lead to habitat destruction, ecosystem service degradation and increased environmental problems such as acid rain. Previous studies have typically used carbon emissions (e.g. Salman et al., 2019; Osobajo et al., 2020; Alaganthiran & Anaba, 2022), greenhouse gas emissions (e.g. Yang et al., 2017; Vasylieva et al., 2019; Sarkodie & Strezov, 2019) or the environmental performance index (e.g. Ave, 2010; Fakher & Abedi, 2017) to investigate the impacts of economic growth on the environment. However, this study focuses on the impacts of economic growth on the ecosystem beyond environmental degradation. Nevertheless, the results of the study are consistent with the literature. The findings are consistent with Bozkurt and Okumus (2017), Al-muali (2012) and Al-muali and Sheau-Ting (2014) who find that energy consumption increases CO2 emissions. Moreover, the results of this study are also consistent with the findings of Mahmood et al. (2019) and Iskenderoglu et al. (2023), which indicate that growth and trade liberalization increase emissions. Similarly, the results of this study are consistent with Farhani et al. (2014) and Bozkurt and Okumus (2017) who find that energy consumption increases environmental damage.

## POLICY IMPLICATIONS

The findings of the study reveal that the impacts of economic growth on ecosystems are complex and vary across ecosystem components. While growth has positive impacts on agriculture and water resources, it clearly has negative impacts on biodiversity, ecosystem services, fisheries and overall ecosystem vitality. Moreover, energy intensity, fossil fuel use and population growth increase environmental pressures, while trade activities also contribute to environmental degradation. These findings make it clear that policies need to be developed to balance the environmental costs of economic growth. Therefore, a number of policy recommendations have been developed to protect the environment and offset the negative impacts of economic growth on the environment: (1) Fossil fuel dependency should be reduced to reduce the negative impact of energy intensity on the environment. Tax incentives and subsidies for the use of renewable energy contribute to lowering energy intensity and reducing environmental damage. (2) Energy efficiency strategies should be implemented to reduce energy intensity. Invest in and deploy energy efficiency technologies in industry, construction, agriculture and transportation sectors. Government support for energy efficiency projects in the industrial sector is critical. (3) Sustainable urban planning should be implemented to reduce the negative impacts of population growth on the environment. Preserving green spaces, building energy-efficient structures and expanding public transportation will reduce pressure on nature. Smart city applications and urban agriculture projects should be promoted to protect rural areas and control urban expansion. (4) Develop sustainable trade policies to reduce the environmental impacts of trade. Adopt green logistics solutions in transport and accelerate the transition to low-carbon transport systems. Low emission technologies in maritime and road transportation should be scaled up and government support should be provided. Encourage policies that reduce carbon footprint and the use of environmentally friendly products. (5) Expand and protect nature reserves to prevent biodiversity loss. Regulations should be made to reduce the environmental impacts of agricultural, industrial and construction activities and environmentally friendly production methods should be encouraged. Sustainable use of natural resources and the establishment of protected areas are essential for sustaining ecosystems. (6) Policies to address climate change should be strengthened to reduce acid rain and other environmental damage. Solutions such as emissions trading and carbon tax can be implemented to reduce carbon emissions. Carbon capture and storage technologies should be encouraged and their use should be expanded. Transition to clean technologies in industry and energy sectors should be supported. (7) Sustainable management of water resources is important for economic growth. Water saving technologies should be promoted, agricultural irrigation should be modernized and water management infrastructure should be strengthened. Wastewater treatment systems should be expanded and environmental pressures on water resources should be reduced. (8) Education and awareness-raising programs should be implemented to raise awareness on environmental sustainability. Individuals and enterprises should be informed about behaviors that will contribute to the environment. Adopting environmentally friendly consumption patterns and promoting sustainable development principles can alleviate environmental pressures. These policy recommendations can help balance economic growth with environmental sustainability and ensure a more livable environment for future generations.

## CONCLUSION

This study investigates the impacts of economic growth on ecosystem in Türkiye for the period 1995-2021. The analysis using the ARDL model reveals the short and long-run impacts of these variables on environmental indicators. The results show that economic growth has a significant impact in all models analyzed. However, the direction of this impact differs according to ecosystem components. Economic growth has a positive impact on agriculture and water resources and a negative impact on biodiversity and habitat, ecosystem services, fisheries, acid rain and total ecosystem vitality. Economic growth has particularly detrimental consequences for biodiversity, ecosystem services, fisheries and overall ecosystem vitality. This suggests that the environmental costs of growth processes need to be considered. Environmentally friendly policies should be combined with sustainable development strategies to reduce the negative impacts of economic growth.

Energy intensity has a negative impact on environmental indicators in almost all models. This result shows that economic growth based on fossil fuel use threatens habitats, water resources and overall ecosystem health by increasing environmentally damaging carbon emissions. Renewable energy and energy efficiency should be promoted to reduce the environmental costs of fossil fuels. Population growth has a negative impact on environmental indicators in most models. Increasing population puts pressure on environmental resources through energy consumption, waste generation and agricultural expansion. Rapid population growth can lead to a reduction in natural habitats and degradation of ecosystems. Therefore, it is critical to adopt sustainable urbanization, agriculture and energy policies to mitigate the environmental impacts of population growth. Trade generally has a negative impact on environmental indicators. Increased trade can increase transportation activities, leading to higher carbon emissions and increased negative impacts on the environment. Green logistics solutions, sustainable production and environmental regulations should be implemented to reduce the environmental impacts of trade.

Although this study reveals important findings, it has some limitations. Future research can address these limitations. First, our study only covers certain environmental and economic indicators. Future studies can expand the dataset. Second, the study is based on a sample of Türkiye. Future research could extend the analysis to emphasize differences between developed and developing countries. Such a comparison could help us comprehend how environmental impacts vary based on the level of economic development. Finally, larger data sets and different analysis techniques could be used to further consolidate and extend the findings of this study.

**Declaration of interests**
The author declares that there is no conflict of interest.

**Author Contributions**
**Emre Akusta:** Conceptualization; data curation; formal analysis; investigation; methodology; software; writing—original draft; writing—review and editing.

## ORCID
**Emre Akusta**  http://orcid.org/0000-0002-6147-5443

## REFERENCES

Agboola, M. O., Bekun, F. V., & Joshua, U. (2021). Pathway to environmental sustainability: nexus between economic growth, energy consumption, CO2 emission, oil rent and total natural resources rent in Saudi Arabia. *Resources Policy*, *74*, 102380. https://doi.org/10.1016/j.resourpol.2021.102380

Agenor, P. R. (2004). Does globalization hurt the poor?. *International economics and economic policy*, *1*, 21-51. https://doi.org/10.1007/s10368-003-0004-3

Alaganthiran, J. R., & Anaba, M. I. (2022). The effects of economic growth on carbon dioxide emissions in selected Sub-Saharan African (SSA) countries. *Heliyon*, *8*(11). https://doi.org/10.1016/j.heliyon.2022.e11193

Al-mulali, U. (2012). Factors affecting CO2 emission in the Middle East: A panel data analysis. *Energy*, *44*(1), 564-569. https://doi.org/10.1016/j.energy.2012.05.045

Al-Mulali, U., & Sheau-Ting, L. (2014). Econometric analysis of trade, exports, imports, energy consumption and CO2 emission in six regions. *Renewable and Sustainable Energy Reviews*, *33*, 484-498. https://doi.org/10.1016/j.rser.2014.02.010

Alptekin, N. (2015). Ranking of EU countries and Turkey in terms of sustainable development indicators: An integrated approach using entropy and TOPSIS methods. *The 9th International Days of Statistics and Economics, Prague, September*, *1012*.

Ansuategi, A., & Escapa, M. (2002). Economic growth and greenhouse gas emissions. *Ecological Economics*, *40*(1), 23-37. https://doi.org/10.1016/S0921-8009(01)00272-5

Antweiler, W., Copeland, B. R., & Taylor, M. S. (2001). Is free trade good for the environment?. *American economic review*, *91*(4), 877-908. https://doi.org/10.1257/aer.91.4.877

Ave, P., & Babolsar, I. (2010). Environmental Performance Index and economic growth: evidence from some developing countries. *Australian journal of basic and applied sciences*, *4*(8), 3098-3102.

Baek, J., Cho, Y., & Koo, W. W. (2009). The environmental consequences of globalization: A country-specific time-series analysis. *Ecological economics*, *68*(8-9), 2255-2264. https://doi.org/10.1016/j.ecolecon.2009.02.021

Bek, N. (2019). Çevresel performans endeksi ve sürdürülebilir yönetişim göstergeleri kapsamında ülke karşılaştırması: Türkiye ve İsviçre örneği. *International Journal of Innovative Approaches in Social Sciences*, *3*(2), 36-45. https://doi.org/10.29329/ijiasos.2019.216.1

Bilen, K., Ozyurt, O., Bakırcı, K., Karslı, S., Erdogan, S., Yılmaz, M., & Comaklı, O. (2008). Energy production, consumption, and environmental pollution for sustainable development: A case study in Turkey. *Renewable and Sustainable Energy Reviews*, *12*(6), 1529-1561. https://doi.org/10.1016/j.rser.2007.03.003

Bozkurt, C., & Okumus, İ. (2017). Gelişmiş Ülkelerde Çevresel Kuznets Eğrisi Hipotezinin Test Edilmesi: Kyoto Protokolünün Rolü. *İşletme ve İktisat Çalışmaları Dergisi*, *5*(4), 57-67.

Copeland, B. R., & Taylor, M. S. (1994). North-South Trade and the Environment. *The Quarterly Journal of Economics*, 109(3), 755–787. https://doi.org/10.2307/2118421

Dickey, D. A., & Fuller, W. A. (1979). Distribution of the estimators for autoregressive time series with a unit root. *Journal of the American statistical association*, *74*(366a), 427-431. https://doi.org/10.1080/01621459.1979.10482531

Dogan, E., Seker, F., & Bulbul, S. (2017). Investigating the impacts of energy consumption, real GDP, tourism and trade on CO2 emissions by accounting for cross-sectional dependence: a panel study of OECD countries. *Current Issues in Tourism*, *20*(16), 1701-1719. https://doi.org/10.1080/13683500.2015.1119103

Engle, R. F., & Granger, C. W. (1987). Co-integration and error correction: representation, estimation, and testing. *Econometrica: journal of the Econometric Society*, 251-276. https://doi.org/10.2307/1913236

Fakher, H. A., & Abedi, Z. (2017). Relationship between environmental quality and economic growth in developing countries (based on environmental performance index). *Environmental Energy and Economic Research*, *1*(3), 299-310. https://doi.org/10.22097/eeer.2017.86464.1001

Farhani, S., Chaibi, A., & Rault, C. (2014). CO2 emissions, output, energy consumption, and trade in Tunisia. *Economic Modelling*, *38*, 426-434. https://doi.org/10.1016/j.econmod.2014.01.025

Gao, J., & Tian, M. (2016). Analysis of over-consumption of natural resources and the ecological trade deficit in China based on ecological footprints. *Ecological indicators*, *61*, 899-904. https://doi.org/10.1016/j.ecolind.2015.10.044

Ghanem, S. K. (2018). The relationship between population and the environment and its impact on sustainable development in Egypt using a multi-equation model. *Environment, development and sustainability*, *20*, 305-342. https://doi.org/10.1007/s10668-016-9882-8

Grossman, G. M., & Krueger, A. B. (1993). 2 Environmental Impacts of a North American Free Trade Agreement. *The Mexico-US Free Trade Agreement*, *11*(2), 13.

Iskenderoglu, Ö., Unlubulduk, S. N., & Karadeniz, E. (2022). Karbon Salınımının Belirleyicileri: Çevresel Performans Endeksi'ndeki Ülkelerde Bir Araştırma. *Verimlilik Dergisi*, 23-36. https://doi.org/10.51551/verimlilik.1058125

Jacobs, C., B., Lee, C., O'Toole, D., & Vines, K. (2014). Integrated regional vulnerability assessment of government services to climate change. *International Journal of Climate Change Strategies and Management*, *6*(3), 272-295. https://doi.org/10.1108/IJCCSM-12-2012-0071

Jalil, A., & Mahmud, S. F. (2009). Environment Kuznets curve for CO2 emissions: a cointegration analysis for China. *Energy policy*, *37*(12), 5167-5172. https://doi.org/10.1016/j.enpol.2009.07.044

Jayanthakumaran, K., Verma, R., & Liu, Y. (2012). CO2 emissions, energy consumption, trade and income: a comparative analysis of China and India. *Energy Policy*, *42*, 450-460. https://doi.org/10.1016/j.enpol.2011.12.010

Johansen, S. (1988). Statistical analysis of cointegration vectors. *Journal of economic dynamics and control*, *12*(2-3), 231-254. https://doi.org/10.1016/0165-1889(88)90041-3

Johansen, S., & Juselius, K. (1990). Maximum likelihood estimation and inference on cointegration—with appucations to the demand for money. *Oxford Bulletin of Economics and statistics*, *52*(2), 169-210. https://doi.org/10.1111/j.1468-0084.1990.mp52002003.x

Karaman, Y. E. (2018). Çevre performans endeksi kapsamında Avrupa Birliği ve Türkiye'nin karşılaştırılması. *Sosyal ve Beşeri Bilimler Dergisi*, *10*(1), 76-85.

Kasman, A., & Duman, Y. S. (2015). CO2 emissions, economic growth, energy consumption, trade and urbanization in new EU member and candidate countries: a panel data analysis. *Economic modelling*, *44*, 97-103. https://doi.org/10.1016/j.econmod.2014.10.022

Kukla-Gryz, A. (2009). Economic growth, international trade and air pollution: A decomposition analysis. *Ecological economics*, *68*(5), 1329-1339. https://doi.org/10.1016/j.ecolecon.2008.09.005

Lopez, R. (1994), "The environment as a factor of production: The effects of economic growth and trade liberalization", *Journal of Environmental Economics and Management*, 27: 163-184. https://doi.org/10.1006/jeem.1994.1032

Madaleno, M., & Nogueira, M. C. (2023). How renewable energy and CO2 emissions contribute to economic growth, and sustainability—an extensive analysis. *Sustainability*, *15*(5), 4089. https://doi.org/10.3390/su15054089

Mahmood, H., Maalel, N., & Zarrad, O. (2019). Trade openness and CO2 emissions: Evidence from Tunisia. *Sustainability*, *11*(12), 3295. https://doi.org/10.3390/su11123295

Nordhaus, W. D. (2010). Economic aspects of global warming in a post-Copenhagen environment. *Proceedings of the National Academy of Sciences*, *107*(26), 11721-11726. https://doi.org/10.1073/pnas.1005985107

Osobajo, O. A., Otitoju, A., Otitoju, M. A., & Oke, A. (2020). The impact of energy consumption and economic growth on carbon dioxide emissions. *Sustainability*, *12*(19), 7965. https://doi.org/10.3390/su12197965

Ozkan, M., & Ozcan, A. (2018). Veri zarflama analizi (VZA) ile seçilmiş çevresel göstergeler üzerinden bir değerlendirme: OECD performans incelemesi. *Yönetim Bilimleri Dergisi*, *16*(32), 485-508.

Perron, P. (1988). Trends and random walks in macroeconomic time series: Further evidence from a new approach. *Journal of economic dynamics and control*, *12*(2-3), 297-332. https://doi.org/10.1016/0165-1889(88)90043-7

Pesaran, M. H., Shin, Y., & Smith, R. J. (2001). Bounds testing approaches to the analysis of level relationships. *Journal of applied econometrics*, *16*(3), 289-326. https://doi.org/10.1002/jae.616

Phillips, P. C., & Perron, P. (1988). Testing for a unit root in time series regression. *Biometrika*, 75(2), 335-346. https://doi.org/10.1093/biomet/75.2.335

Pimonenko, T. V., Liulov, O. V. and Chyhryn, O. Y. (2018). Environmental Performance Index: relation between social and economic welfare of the countries. *Environ. Econ.*, 9, 1-11. https://doi.org/10.21511/ee.09(3).2018.01

Pinar, M. (2022). Sensitivity of environmental performance index based on stochastic dominance. *Journal of Environmental Management*, *310*, 114767. https://doi.org/10.1016/j.jenvman.2022.114767

Rehman, A., Radulescu, M., Ma, H., Dagar, V., Hussain, I., & Khan, M. K. (2021). The impact of globalization, energy use, and trade on ecological footprint in Pakistan: does environmental sustainability exist?. *Energies*, *14*(17), 5234. https://doi.org/10.3390/en14175234

Richmond, A. K., & Kaufmann, R. K. (2006). Is there a turning point in the relationship between income and energy use and/or carbon emissions?. *Ecological economics*, *56*(2), 176-189. https://doi.org/10.1016/j.ecolecon.2005.01.011

Rock, M. T. (1996). Pollution intensity of GDP and trade policy: can the World Bank be wrong?. *World development*, *24*(3), 471-479. https://doi.org/10.1016/0305-750X(95)00152-3

Salman, M., Long, X., Dauda, L., & Mensah, C. N. (2019). The impact of institutional quality on economic growth and carbon emissions: Evidence from Indonesia, South Korea and Thailand. *Journal of Cleaner Production*, *241*, 118331. https://doi.org/10.1016/j.jclepro.2019.118331

Sarkodie, S. A., & Strezov, V. (2019). Effect of foreign direct investments, economic development and energy consumption on greenhouse gas emissions in developing countries. *Science of the total environment*, *646*, 862-871. https://doi.org/10.1016/j.scitotenv.2018.07.365

Shahbaz, M., Hye, Q. M. A., Tiwari, A. K., & Leitão, N. C. (2013). Economic growth, energy consumption, financial development, international trade and CO2 emissions in Indonesia. *Renewable and sustainable energy reviews*, *25*, 109-121. https://doi.org/10.1016/j.rser.2013.04.009

Tabash, M. I., Farooq, U., Aljughaiman, A. A., Wong, W. K., & AsadUllah, M. (2024). Does economic complexity help in achieving environmental sustainability? New empirical evidence from N-11 countries. *Heliyon*, *10*(11). https://doi.org/10.1016/j.heliyon.2024.e31794

Turhan, E., Cerit Mazlum, S., Şahin, Ü., Şorman, A. H., & Cem Gündoğan, A. (2016). Beyond special circumstances: climate change policy in Turkey 1992–2015. *Wiley Interdisciplinary Reviews: Climate Change*, *7*(3), 448-460. https://doi.org/10.1002/wcc.390

Ulucak, R., & Erdem, E. (2012). Çevre-iktisat ilişkisi ve Türkiye'de çevre politikalarinin etkinliği. *Akademik Araştırmalar ve Çalışmalar Dergisi (AKAD)*, *4*(6), 78-98.

Usman, M., Kousar, R., Yaseen, M. R., & Makhdum, M. S. A. (2020). An empirical nexus between economic growth, energy utilization, trade policy, and ecological footprint: a continent-wise comparison in upper-middle-income countries. *Environmental Science and Pollution Research*, *27*, 38995-39018. https://doi.org/10.1007/s11356-020-09772-3

Van Dao, N., & Van, V. H. (2020). Population explosion and the environment in developing countries: A case study of Vietnam. *Revista Argentina De Clinica Psicológica*, *29*(4), 202.

Vasylieva, T., Lyulyov, O., Bilan, Y., & Streimikiene, D. (2019). Sustainable economic development and greenhouse gas emissions: The dynamic impact of renewable energy consumption, GDP, and corruption. *Energies*, *12*(17), 3289. https://doi.org/10.3390/en12173289

Wolf, M. J., Emerson, J. W., Esty, D. C., de Sherbinin, A., Wendling, Z. A., *et al*. (2022). *2022 Environmental Performance Index*. New Haven, CT: Yale Center for Environmental Law & Policy. epi.yale.edu

Yang, X., Lou, F., Sun, M., Wang, R., & Wang, Y. (2017). Study of the relationship between greenhouse gas emissions and the economic growth of Russia based on the Environmental Kuznets Curve. *Applied energy*, *193*, 162-173. https://doi.org/10.1016/j.apenergy.2017.02.034

Yilanci, V., Pata, U. K., & Cutcu, I. (2022). Testing the persistence of shocks on ecological footprint and sub-accounts: evidence from the big ten emerging markets. *International Journal of Environmental Research*, *16*(1), 10. https://doi.org/10.1007/s41742-021-00391-5